\documentclass[tradiabstract]{aa}
\usepackage{graphicx}
\usepackage{natbib}
\usepackage{txfonts}
\usepackage{color}

\newcommand{\HI}{{\ion{H}{1}}}

\newcommand{\kms}{$\,$km$\,$s$^{-1}$}
\newcommand{\ergs}{$\,$erg$\,$s$^{-1}$}

\newcommand{\mJybeam}{mJy beam$^{-1}$}
\newcommand{\muJybeam}{$\mu$Jy beam$^{-1}$}
\newcommand{\mJybeamch}{mJy beam$^{-1}$ ch$^{-1}$}
\newcommand{\msun}{{${M}_\odot$}}
\newcommand{\msunyr}{{${ M}_\odot$ yr$^{-1}$}}

\newcommand{\tspin}{$T_{\rm spin}$}

\def\HI{H{\,\small I}}
\def\Na{Na{\,\small I}}

\def\emph#1{{\sl #1}}
\newcommand{\ltsima} {$\; \buildrel < \over \sim \;$}
\newcommand{\gtsima} {$\; \buildrel > \over \sim \;$}
\newcommand{\lta} {\lower.5ex\hbox{\ltsima}}
\newcommand{\gta} {\lower.5ex\hbox{\gtsima}}

\newcommand{\miriad}{{{MIRIAD}}}

\begin{document}

%\copyright~\textsf{\copyright}

%
\title{Another piece of the puzzle: the fast \HI\ outflow in Mrk~231}
\titlerunning{Fast \HI\ outflow in Mrk~231}
\authorrunning{Morganti et al.}
\author{Raffaella Morganti\inst{1,2},  Sylvain Veilleux\inst{3}, Tom Oosterloo\inst{1,2}, Stacy H. Teng\inst{4}, David Rupke\inst{5}}

\institute{Netherlands Institute for Radio Astronomy, Postbus 2,
7990 AA, Dwingeloo, The Netherlands
\and
Kapteyn Astronomical Institute, University of Groningen, Postbus 800,
9700 AV Groningen, The Netherlands
\and
Department of Astronomy, Joint Space-Science Institute, University of Maryland, College Park, MD 20742, USA
\and
Science and Technology Division, Institute for Defense Analyses, Alexandria, VA 22311, USA
\and
Department of Physics, Rhodes College, Memphis, TN 38112, USA
}
\offprints{morganti@astron.nl}

\date{Received ...; accepted ...}

\date{\today}

\abstract{
We present the detection, performed with the Westerbork Synthesis Radio Telescope (WSRT) and the Karl Jansky Very Large Array (VLA), of a fast  \HI\ 21cm outflow  in the ultra-luminous infrared galaxy Mrk~231.   The outflow is observed as  shallow \HI\ absorption blueshifted $\sim$1300 \kms\ with  respect to the systemic velocity  and located against the inner kpc of the radio source. The outflowing gas has an estimated  column density between 5  and $15 \times 10^{18} T_{\rm spin}$ cm$^{-2}$. We derive the \tspin\  to lie in the range  400 -- 2000 K and the corresponding \HI\ densities are  $n_{\rm HI} \sim 10 - 100$ cm$^{-3}$.
Our results  complement   previous findings and confirm the multiphase nature of the outflow in Mrk~231.   Although effects of the interaction between the radio plasma and the surrounding medium cannot be  ruled out, the energetics and the lack of a clear kpc-scale jet  suggest that the most likely origin of the \HI\ outflow is a wide-angle nuclear wind, as earlier proposed to explain the neutral outflow traced by \Na\ and molecular gas in this source. 
Our results suggest that an \HI\ component   is present in fast outflows regardless  of the acceleration mechanism (wind {\sl vs} jet driven) and that it must be connected with common properties of the pre-interaction gas involved.
Considering the observed similarity  of their column densities, the \HI\ outflow  likely represents the inner part of the broad wind  identified on larger scales in atomic \Na. 
The mass outflow rate of the \HI\ outflow (between 8 and 18 \msunyr) does not appear to be as large as the one observed in molecular gas, partly due to the smaller sizes of the outflowing region sampled by the \HI\ absorption. These characteristics are commonly seen in other cases of AGN-driven outflows suggesting that the \HI\ may  represent a short intermediate  phase in the rapid cooling of the gas. 
The results further confirm  \HI\  as a good tracer for AGN-driven outflows not only in powerful radio sources.  
We also obtained deeper continuum images than previously available.
They confirm the complex structure of the radio continuum originating both from the AGN and
star formation. At the resolution obtained with the VLA ($\sim$$1^{\prime\prime}$) we do not see a kpc-scale  jet. Instead, we detect a plateau of emission, likely due to star formation, surrounding the  bright nuclear region. We also detect a poorly collimated bridge  which may represent the channel feeding the  southern lobe. The unprecedented depth of the low-resolution
WSRT image reveals radio emission extending 50$^{\prime\prime}$ (43 kpc) to the south
and 20$^{\prime\prime}$ (17 kpc) to the north.}

\maketitle  

\section{Cold gas and fast, massive outflows}
\label{sec:introduction}

AGN-driven outflows have recently attracted considerable attention for their potential impact on galaxy evolution because they may play an important role in regulating star formation as well as the growth of the central super-massive black hole   (SMBH; \citealt{Veilleux05,Bland07,Fabian13}). Understanding their occurrence, origin and physical characteristics is key for quantifying their impact.  Gas outflows are now recognised as being multi-phase and a large body of literature is available on this topic, including studies of the hot and warm ionised gas (e.g., \citealt{Nesvadba08,Reeves09,Holt09,Harrison12,Harrison14,Tombesi15}),  the atomic gas (e.g., \citealt{Rupke11,Rupke13a,Rupke15,Lehnert11,Morganti13,Morganti15}), warm and cold molecular gas (e.g., \citealt{Feruglio10,Dasyra11,Guillard12,Rupke13b,Garcia2014,Tadhunter14,Cicone14})  as well as  OH (e.g., \citealt{Fischer10,Sturm11,Veilleux13}).
Considering the large amounts of energy released by the AGN,  one of the main open questions is the presence of  large amounts of cold gas  participating in these multi-phase AGN-driven outflows.

%============================ Panel literature radio continuum ==================

\begin{figure*}
	\centering
	\includegraphics[width=18.5cm,angle=0]{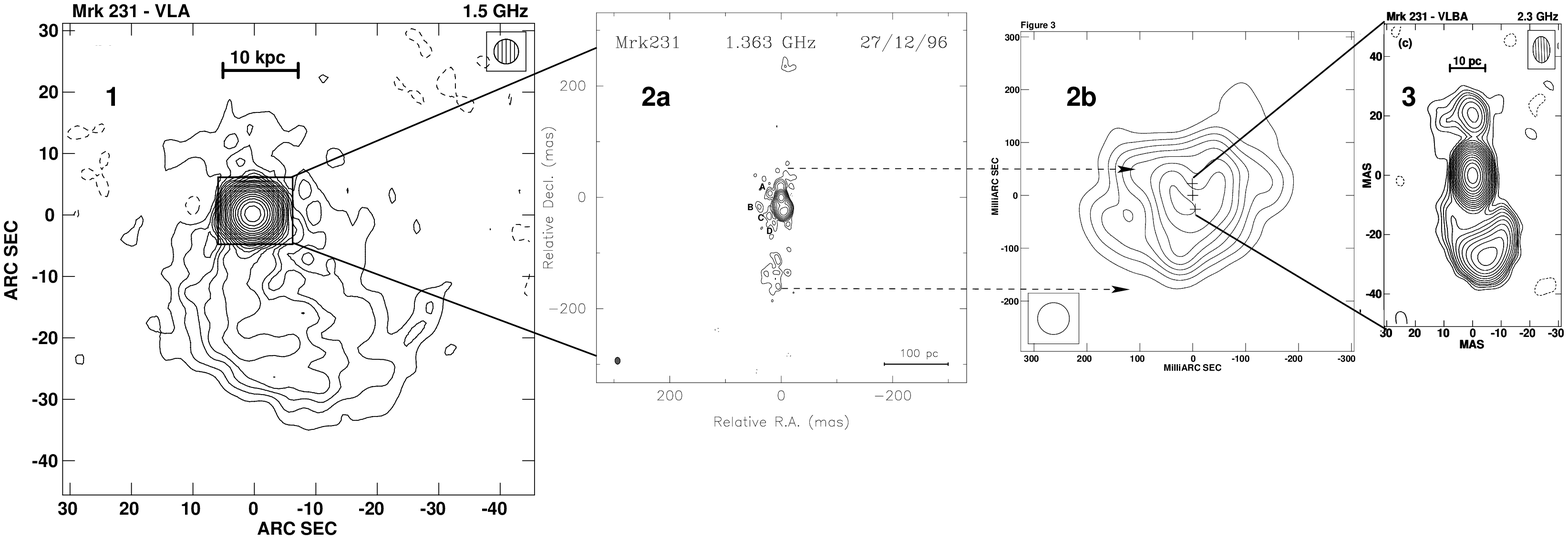}
	\caption{\label{fig:contPanel} Overview of the radio continuum structures  of Mrk~231 ranging from tens of kpc to the inner tens of pc regions (at the distance of Mrk~231 $1^{\prime\prime}$ corresponds to 0.867 kpc). The images, taken from \citet{Ulvestad99,Carilli98,Taylor99} (\copyright AAS, reproduced with permission), show the presence of different structures, see text for details. The disk-like structure (Panels 2b)  aligned almost E-W has been detected after the subtraction of the brighter nuclear structure (Panel 3). For details, see description in the text. 	}
\end{figure*}

%========================================================================

Different mechanisms have been proposed to accelerate the gas. The most widely considered are wide-angle, wind-driven outflows, launched from the accretion disk and driven by radiation pressure or by a hot thermal wind (see \citealt{Veilleux05} for an overview), and outflows driven by the  mechanical action of the radio plasma emanating from the AGN \citep{Wagner11,Wagner12}.  
Connected to these different mechanisms are  questions about the physical conditions of the gas resulting from each of them, which phases of the gas  can be best  used as tracers, where do the outflows occur with respect to the central SMBH and what is their contribution to the energetics in general. Interestingly, some of the theoretical models are now trying to account for the presence of a cold component in the fast outflowing gas \citep{Mellema02,Faucher12,Zubovas14,Costa14,Nims15}. 

In this context, detailed studies of single objects are essential in order to more precisely characterize the
gaseous outflow. They can provide the relevant parameters (such as the location and extent of the outflow) and  physical properties  (like temperature, density, mass, mass outflow rate) which can be used for a comparison with  theoretical models.  
The ultra-luminous infrared galaxy Mrk~231 is an ideal object for such a study.  This galaxy is  an ongoing merger where a starburst has  started recently and an AGN has also been triggered. This is an interesting, albeit likely short-lasting, phase in the evolution of a galaxy. Given these conditions, it is not surprising that Mrk~231 represents one of the best cases where an AGN-driven outflow is observed in many different gas phases (see  \citealt{Feruglio15}  for a recent summary)
and over a wide range of scales (e.g., \citealt{Veilleux16}). This last paper also argues that Mrk 231 is the nearest example of
weak-lined wind-dominated quasars with high Eddington ratios and geometrically thick ("slim") accretion disks.

Among the various manifestations of the active nucleus, Mrk~231 has also a radio source. 
This allows to investigate the characteristics and kinematics of the neutral hydrogen (\HI\ 21 cm) in the nuclear regions by observing this gas in absorption against the radio continuum. 
Although \HI\ absorption was already detected in Mrk~231 early on \citep{Carilli98}, only recently an extra broad and blueshifted component, likely tracing the outflow of atomic neutral gas, was discovered in observations done with the  Westerbork Synthesis Radio Telescope \citep[WSRT;][]{Morganti11}  and the Green Bank Telescope  \citep[GBT;][]{Teng13}. 

Here we present additional WSRT observations (with a spatial resolution of about 10$^{\prime\prime}$), complementing those originally showing the broad, blueshifted absorption, as well as follow up observations with the Karl G.\ Jansky Very Large Array (VLA)  tracing the \HI\ outflow at higher spatial  resolution ($\sim$$1^{\prime\prime}$). Our main goal is to determine the location of the outflow of atomic hydrogen in order to relate it to the other phases of the gas, but the data also provide  interesting information about the structure of the radio continuum of Mrk~231.

\section{Overview of the known radio properties of Mrk~231}

We start with a summary of what is known about the complex structure of the radio continuum  and \HI\ absorption in Mrk~231.
Mrk~231 is a famous and well studied ultra-luminous infrared galaxy (ULIRG, log[$L_{\rm IR}/L_{\odot}] = 12.37$), often referred to as the closest quasar (QSO)\footnote{At the distance of Mrk~231 ($z=0.0422$, V$_{\rm sys} =12642$ \kms) $1^{\prime\prime}$ corresponds to 0.867 kpc)}. The optical morphology clearly shows that Mrk~231 is an ongoing major merger (e.g., \citealt{Veilleux02,Veilleux06}). The object hosts an AGN  as well as a young, dusty starburst characterised by a star formation rate (SFR) of $\sim$$ 160$ $M_{\odot}$ yr$^{-1}$  (see \citealt{Veilleux09} and Table 1 in \citealt{Rupke13a}). 
The radio power of Mrk~231 is log $P_{\rm 1.4\ GHz} = 24.15$ W Hz$^{-1}$, corresponding to the top of the distribution for Seyfert galaxies (i.e.\ brighter than e.g.\ NGC~1068 and IC~5063) and to the lower end of the distribution for radio galaxies. Below we summarise the known radio characteristics (continuum and \HI) relevant for the study presented in this paper.

%============================ WSRT absorption  ======================================

\begin{figure}
	\centering
	\includegraphics[width=9cm,angle=0]{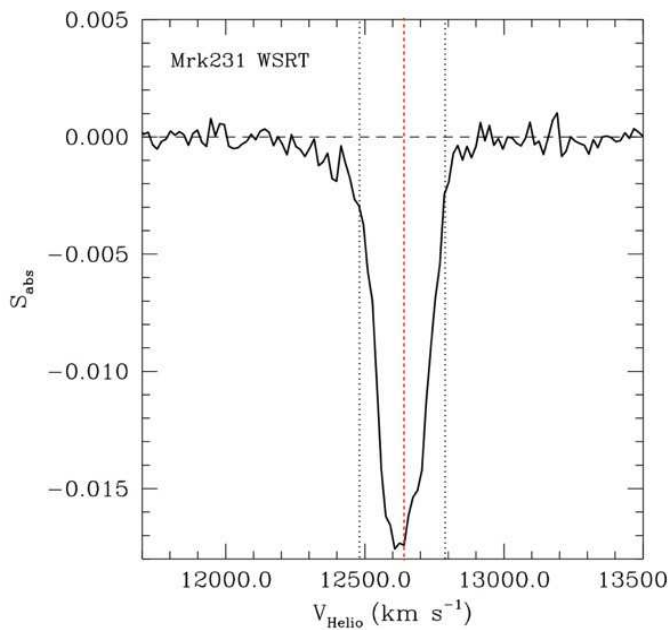}
	\caption{\label{fig:HIprofWSRT} \HI\ absorption profile from the WSRT observations ($1^{\prime\prime} =  0.867$ kpc). The red line represents the systemic velocity while the black dotted lines represent the range of absorption detected by \citet{Carilli98} and identified with the rotating circumnuclear disk. }
\end{figure}
%===========================================================================================

\subsection{Radio continuum}
\label{sec:literatureContinuum}

The radio source in Mrk~231 has a complex morphology \citep{Baum93,Ulvestad99,Carilli98,Taylor99}. It includes different structures on scales  from pc to tens of kpc. The morphology of the radio continuum is characterised by three main structures (see Fig.\ \ref{fig:contPanel}): the southern extended lobe (labelled 1 in Fig.\ \ref{fig:contPanel}),  structures of sub-kpc scale emitted in the north-south direction (labelled 2a and 3 in Fig.\ \ref{fig:contPanel}) and  disk-like emission extending $\sim$200 mas from the nucleus in the east-west direction (labelled 2b in Fig.\ \ref{fig:contPanel}).

The large-scale lobe structure is highly asymmetric. It extends about 30~kpc in the south direction,  although more extended emission was found  by \citet{Baum93} mostly to the south-east,  and shows a distorted morphology with a bending to the west. Only a barely visible extension is observed to the north direction (see Fig. \ref{fig:contPanel} panel 1).

The  north-south  structures on  sub-kpc scale are the brighter components. The studies of \citet{Carilli98} and \citet{Taylor99} show that they have a morphology resembling  "bubbles", suggesting that the radio plasma is perhaps ejected via discrete events in the north-south direction. These structures are reminiscent of the poorly collimated radio emission often observed  in  Seyfert galaxies (e.g.\ Morganti et al.\ 1999 and refs therein). Those structures are characterised by low velocity of the radio plasma and, therefore, are dominated by turbulence resulting in the entrainment of a large thermal component which tends to dominate the flow (Bicknell et al.\ 1998).  The absence of prominent free-free opacity \citep{Taylor99} in these radio components indicates that at least some parts of the nuclear region have a relatively unobstructed line of sight to the observer, perhaps cleared out by the AGN. 

The east-west  disk (Panel 2b in Fig. \ref{fig:contPanel}) has a spectral index consistent with non-thermal synchrotron emission from a population of relativistic electrons accelerated in shocks driven by supernova remnants \citep{Taylor99}.
Given its gaseous molecular component (\citealt{Downes98} and see below) its emission could be connected to star formation \citep{Taylor99}. 
Interestingly, unlike the north-south structure,  the integrated spectrum of the inner part of the disk ($<100$ mas) shows an inversion below 1.3~GHz, most likely due to free-free absorption \citet{Taylor99}.  
Fitting a free-free absorption model to the data, \citet{Taylor99} obtain an emission measure, EM = $7.9 0.6 \times 10^5$ ($T_{\rm K}/10^4)^{3/2}$ pc cm$^{-6}$, where $T_{\rm K}$ is the kinetic temperature of the gas. Using a disk thickness of 23 pc (derived from CO observations, \citet{Downes98}) they derive  $n_{\rm e} = 185$ ($T_{\rm K}/10^4)^{3/4}$ cm$^{-3}$.
This density is comparable to the density of the  gas associated with the strong \HI\ absorption (see below) obtained assuming a spin temperature of 1000 K ($n_{\rm \HI} \sim 0.3$ \tspin\ cm$^{-3}$ \citep{Carilli98}.

Finally, because of the large variability seen in its nucleus, Mrk~231 has been proposed to be a blazar-type object \citep{Reynolds13,Lindberg15}. According to \citet{Reynolds09}, the radio emission is ejected almost along the line of sight ($\Theta_{\rm max}  < 25.6^{\circ}$ ) and the southern lobe is the one coming toward the observer. 

\subsection{\HI\ absorption}
\label{sec:literatureLine}

A deep \HI\ absorption feature has been detected with the VLA by \citet{Carilli98}. This absorption is centred on the systemic velocity of Mrk~231.
Further VLBA observations have shown that, surprisingly, this absorption is not occurring against the bright central core (Panel 3, in Fig. \ref{fig:contPanel}). Instead, the absorption is seen against the $\sim$200 mas (i.e.\ 170 pc) disk (2b in Fig. \ref{fig:contPanel}). This absorption shows a  velocity gradient  in  the E-W direction across this structure  \citep{Carilli98,Taylor99}.  

As remarked above, this structure is likely the inner part of the disk seen in CO emission by \citet{Downes98} and recently confirmed by \citet{Feruglio15}. The molecular disk (with a position angle of the major axis of $77^{\circ}$)  has a  thickness of 23 pc and must be very close to face-on, with $i < 20^{\circ}$ as derived by \citet{Downes98}. 
Also the study of the radio continuum by \citet{Carilli98} concluded that the disk against which the \HI\ absorption occurs cannot be oriented too far from the sky plane, or else \HI\ absorption should have been detected against the radio core. For the same reason,  the parsec-scale radio lobe must be oriented along the line of sight.

%============================ WSRT radio continuum ======================================

\begin{figure}
	\centering
	\includegraphics[width=9cm,angle=0]{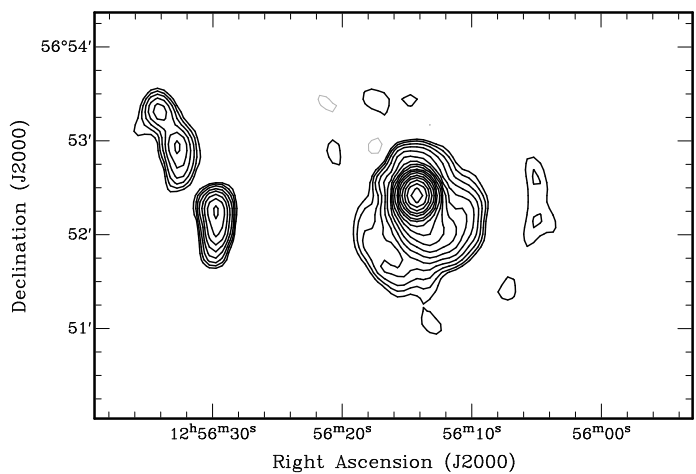}
	\caption{\label{fig:ContWSRT} WSRT continuum image. The contour levels range from -0.3, 0.3 \mJybeam\  to 250 \mJybeam\ with  increasing factors of 1.5. The radio source  E of Mrk~231 is an unrelated background/foreground object.}
\end{figure}
%=====================================================================================

\section{WSRT observations}

We (PI R.\ Morganti, proposals S10B/007 and  R11A/022) observed Mrk~231  with the WSRT at different epochs. The first observations, using service time,  led to the spectrum showing the blueshifted outflow presented in \citet{Morganti11}. Two full synthesis observations were later obtained on April 13 and 15, 2011 (see Table \ref{tab:obs} for details).
We used a 20-MHz bandwidth divided in 1024 channels and centred on 1362.87 MHz.  The data were reduced using the \miriad\ software \citep{Sault95} following standard recipes. Cubes were made with uniform and with a robust weighting   of 0.5  \citep{Briggs95}  resulting in a  noise level of 0.48 \mJybeamch\  and 0.32 \mJybeamch\ respectively  for a  velocity resolution of 16 \kms.  After Hanning smoothing of the data, the noise level reached 0.37 \mJybeamch\  and 0.25 \mJybeamch\  for a velocity resolution of 32 \kms.
The spatial resolution of the cubes is  $12.2^{\prime\prime} \times 10.3^{\prime\prime}$ (P.A. $ = 0.6^{\circ}$) and $21.0 \times 17.0$ (P.A. $ = 0.5^{\circ}$) for the uniform and the robust 0.5 weighting respectively with the position angle measured from north to
east.

Our new observations confirm, in addition to the deep absorption already reported by \citet{Carilli98},  the broad blueshifted wing in the \HI\ profile as shown in Fig.\ \ref{fig:HIprofWSRT} which was seen already in the initial profile from the service observations \citep{Morganti11} and in observations with the GBT  \citep{Teng13}. 

The absorption components (both the deep and the shallow) are detected only against the peak of the radio continuum. However, given the relatively low spatial resolution of the WSRT observations, this corresponds to a region of about 9~kpc. To obtain more information on the location of the \HI\ absorption, we have performed higher resolution follow up observations,  see Sec. \ref{sec:VLAobs}.

We have used the line-free channels to obtain a continuum image, which is shown in Fig.\  \ref{fig:ContWSRT}. The rms noise in this image is 0.085 \mJybeam\ and  the restoring beam is $13.6^{\prime\prime} \times 11.1^{\prime\prime}$ (P.A. = 0.4$^\circ$). The peak of the continuum emission is 253.5 mJy.
The radio continuum structure at this resolution includes a bright central source and a north-south extension which follows the sub-kpc structure shown by \citet{Carilli98} and \citet{Ulvestad99}. 
The large-scale radio structure mapped by the WSRT shows no evidence for highly collimated jets. The southern extended region shows bends to the west, consistent with higher resolution images.
However, the WSRT radio continuum image also shows  more extended features, confirming most of the early large-scale structure, in particular the south-east extension,  seen by \citet{Baum93}. In particular, the extension to the north is now clearly seen while only an hint was visible in the image of \citet{Ulvestad99}. 
Furthermore, the diffuse part of the southern lobe is more extended and, in particular, the WSRT image shows an extension to the east which was not seen before. 

%================= VLA absorption and zoom-in + Gaussian Fit  ===============================

\begin{figure*}
	\centering
{\includegraphics[width=9cm,angle=0]{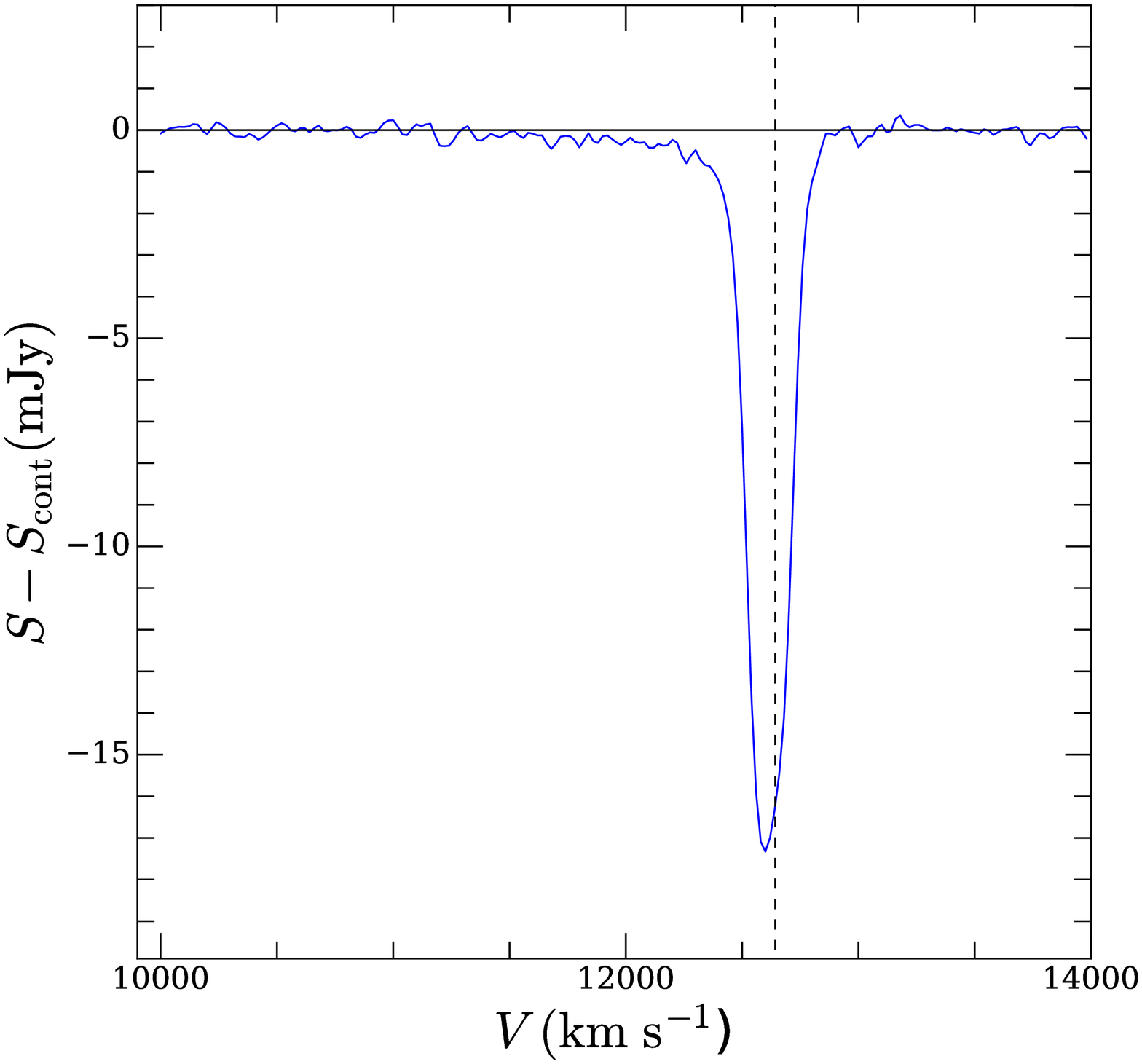}\hfill
\includegraphics[width=9cm,angle=0]{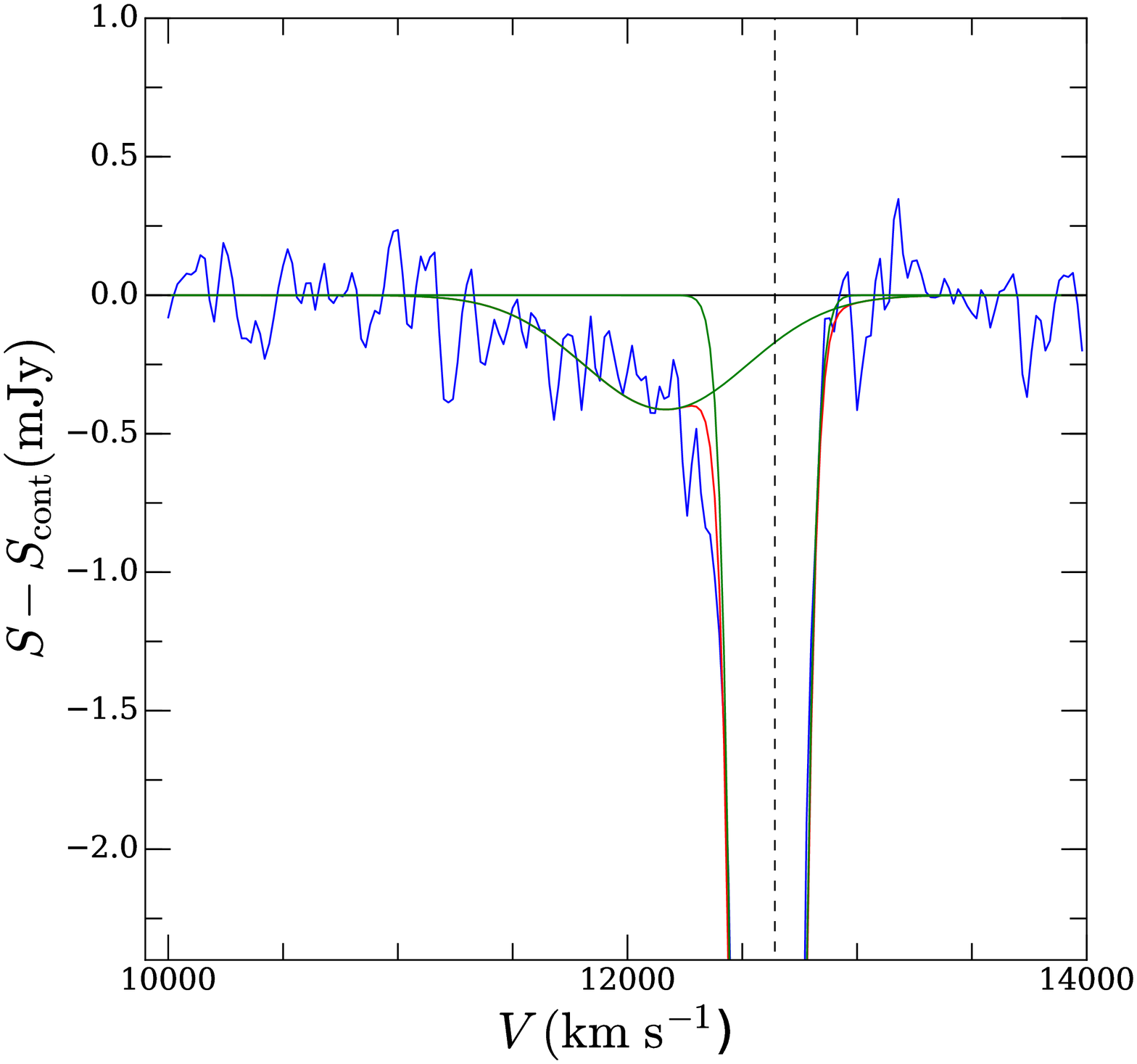}}
	\caption{\label{fig:HIprofJVLA}  {\bf Left:} \HI\ absorption profile from the VLA data. The shallow broad absorption is visible at velocities below $\sim 12250$ \kms. The dashed line indicate the systemic velocity of Mrk 231. 
{\bf Right:}	Zoom-in of the  \HI\ absorption profile from the VLA data. 
	 that better  shows  the blueshifted wing. Superposed is the Gaussian fit of the broad component discussed in the text.}
\end{figure*} 

%===================================================================================

\begin{table} 
\caption{Summary of the observations used in this paper}
\centering 
\begin{tabular}{lcc} 
\hline\hline 
\multicolumn{3}{c}{WSRT}    \\
\hline
 Date        & April 13 \& 15, 2011  & \\
 On source time     & $2\times12$ h   & \\
 Beam$^a$ (unif)     &   $12.2^{\prime\prime} \times 10.3^{\prime\prime}$ ($0.6^{\circ}$)   \\
 Beam$^a$ (r0.5)     &   $21.0^{\prime\prime} \times 17.0^{\prime\prime}$ ($0.5^{\circ}$)   \\
 Peak continuum & 253.5 mJy & \\
 Peak absorption & $ - 18.5$ mJy & \\
 rms noise  line  (unif)   &   0.48 - 0.37\mJybeam\    &                 \\
 rms noise  line  (r0.5)   &   0.32 - 0.25\mJybeam\     &  \\
 rms noise  continuum & 0.053 \mJybeam\ &   \\
\hline
\multicolumn{3}{c}{VLA}    \\
\hline
 Date        & March 15, 27 \& 29, 2014  & \\
On source time      & $3 \times 5$ h  & \\
 Beam$^a$  (unif)    &  $ 0.9^{\prime\prime} \times 0.86^{\prime\prime}$  ($-86^{\circ}$)   \\
 Beam$^a$  (r0.5)    &   $1.4 ^{\prime\prime}\times 1.3^{\prime\prime}$ ($-83^{\circ}$)  &          \\
 Peak continuum & 263 mJy& \\
 Peak absorption & $-19.3$ mJy & \\
 rms noise  line (unif)    &   0.40 - 0.28 \mJybeam\   &                      \\
  rms noise  line (r0.5)    &   0.20 - 0.14 \mJybeam\   &                      \\
  rms noise  continuum (1.3~GHz) & 0.02 \mJybeam\ &   \\
  rms noise  continuum (1.6~GHz) & 0.04 \mJybeam\ &   \\
\hline
\end{tabular}
$^a$ P.A. measured from north to east
\label{tab:obs}
\end{table}

\section{VLA  observations}
\label{sec:VLAobs}

We (PI S.\ Veilleux, proposal \#14A-389) used the VLA in A configuration in order to achieve the highest possible spatial resolution to trace the location of the \HI\ outflow. We obtained a total of 15~h of useful data.  The observations were done with a 64~MHz bandwidth centred on the frequency of the redshifted \HI\ (1362.87 MHz) and using 1024 channels, allowing to obtain a velocity resolution of 13.7 \kms\ (see Table \ref{tab:obs} for details). However, given that we are  interested in the blueshifted wing and that this wing is particularly faint, the final cubes have been made with a channel width of 16 \kms, i.e. the same as for the WSRT cubes, and a subsequent Hanning smoothing applied (i.e.\ with a  final velocity resolution of $\sim$32 \kms) in order to improve the sensitivity.
Observations of 10 minutes on  target were alternated with 2-minute scans on the secondary calibrator J1313+54581.  The flux scale was determined by using 3C~286 assuming a flux at 1.362 GHz of 14.84 Jy. For the bandpass calibration we have used the secondary calibrator which appears unresolved. The total integration time on the secondary was sufficiently large to allow us to use it as bandpass calibrator without increasing the noise in the target data. 

The data calibration and reduction was done using the \miriad\ package (Sault et al.\ 1995) following  standard steps  (self calibration, continuum subtraction, mapping/cleaning). 

The final cubes were made using  uniform weighting and robust weighting of 0.5. The former has the highest spatial resolution ($0.9^{\prime\prime} \times 0.86^{\prime\prime}$, P.A. = $- 86^{\circ}$) corresponding to about $\sim 0.8$ kpc and has a  noise level of $ 0.4$ \mJybeamch\ before and  $  0.28$  \mJybeamch\  after Hanning smoothing. The cube with robust 0.5 weighting has a resolution of  $1.4 \times 1.3$ (P.A. =$-83^{\circ}$) corresponding to about 1.2~kpc. The noise level in the cubes are  0.2  \mJybeamch\   and 0.14  \mJybeamch\  respectively, after Hanning smoothing. 

A continuum image was obtained using the line-free channels of the data.  
We also made continuum images using  uniform and robust 0.5 weighting. In the remainder of the paper we will be using the robust 0.5 image because it better illustrates  the important features of the source. This image has an  rms noise level of 20 \muJybeam.

The VLA data include simultaneous observations in a second band centered on 1606.925 MHz, using a bandwidth of 128~MHz. The flux scale was derived from 3C~286, assumed to be 13.9 Jy at this frequency. This second observing band turned out to be much more affected by radio frequency interference (RFI) and, therefore, did not provide an image reaching the theoretical depth expected from the broader band.  Thus, the quality of this continuum image is actually lower than of the image at 1.3~GHz. The image at 1.6~GHz has an rms noise level of 40 \muJybeam.
Nevertheless, we have used this image to obtain a first-order integrated spectral index of some of the extended regions. 

\begin{table} 
\caption{Parameters of the double-Gaussian used in the fit of the \HI\ absorption profile  shown in Fig.\ \ref{fig:HIprofJVLA} (right).}  
\centering 
\begin{tabular}{l r@{ $\pm$ }l r@{ $\pm$ }l }
\hline\hline
&\multicolumn{2}{c}{component 1}
&\multicolumn{2}{c}{component 2}\\
                    \hline
 Peak (mJy) 							& --17.96 & 0.15   &  --0.41 & 0.06  \\
 Peak velocity (\kms)    	& 12612.8 & 0.7   & 12166.6 & 91.8 \\
 FWHM  (\kms)   					&   197.3 & 2.0   &   839.0 & 185.9       \\
 Integral  (mJy \kms)			& --3780.1 & 49.0 & --369.3 & 97.1\\
\hline
\end{tabular}
\label{tab:GaussFit}
\end{table}

\section{Results}

\subsection{The broad, blueshifted  component of the \HI\  absorption} 
\label{sec:HIabs}

The new observations confirm the presence of a broad, blueshifted component in the \HI\ absorption profile of Mrk~231 as illustrated in  Figs \ref{fig:HIprofWSRT}  and  \ref{fig:HIprofJVLA}.  Because it is observed in absorption, we can unambiguously identify the \HI\ gas to be in front of the radio source and, because it is blueshifted with respect to the systemic velocity, to be part of an outflow.  To first order, the profile obtained from the VLA data is very similar the one obtained with the WSRT.  However, thanks to the higher sensitivity of the VLA data, the blueshifted wing in the VLA profile appears to be broader, reaching velocities down to about 11300 \kms. 

We have performed a fitting of the \HI\ profile to derive the parameters of the  absorption and, in particular, of the broad part. Two Gaussian components were fitted as shown in  Fig.\ \ref{fig:HIprofJVLA} (right) and the parameters of the fits are listed in Table \ref{tab:GaussFit}.
The blueshifted component has a FWHM of 837 \kms\  and reached velocities up to $\sim$1300 \kms\ blueshifted compared to the systemic velocity. The peak of the shallow absorption resulting from the fit is only $\sim$ 0.41  mJy.  Even in the high-resolution  VLA observations, the broad \HI\ absorption is seen only against the peak of the continuum. The amplitude of the shallow absorption  as measured by the VLA is very similar to what is seen in the WSRT profile. Therefore we conclude that  the \HI\ outflow is confined to the inner kpc.  

As expected, we also detect  the deep \HI\ absorption component originally found by  \citet{Carilli98} and which is associated with the inner gas disk. 
Interestingly, the amplitude of this absorption is similar at the very different spatial resolutions probed by the available observations, ranging from WSRT, VLA down to the milliarcsecond resolution of the VLBA data presented by \citet{Carilli98}. This suggests that the $\sim$ 200 mas radio continuum from the disk structure imaged by the VLBA represents the  full extent of the background radio continuum against which this deep component of \HI\ is detected. 

The physical parameters of the \HI\ outflow depend on its location. Given that we do not spatially resolve the outflow and given the complex continuum structure in the inner kpc of Mrk~231 (see Sec.\ \ref{sec:literatureContinuum}), we need to consider at least two extreme scenarios. 
The blueshifted \HI\ component could be part of a flow distributed over a large opening angle. In this scenario the broad wing is due to gas in front of the circumnuclear disk (labelled 2b in Fig.\ \ref{fig:contPanel}) which has a radius $\sim$ 200 mas (170 pc) and a total flux 130 mJy  \citep{Carilli98}).   
A second possibility is that the \HI\ outflow corresponds to gas  pushed out by the interaction with the radio "bubble", i.e.\  the inner north-south VLBA structure (up to distance $\sim$30 pc from the nucleus, labelled  3 in Fig.\ \ref{fig:contPanel}). The absorption would most likely occur against the  southern radio "bubble" (which has a  flux of 44 mJy, \citealt{Carilli98}) or against the "bubble" and the core (with a total flux of these two components of 84 mJy, from \citealt{Carilli98}).  

Following these two extreme situations, we estimate that the optical depth $\tau$ of the \HI\ (defined as $\tau = \rm{ln}[1-S/(S_{\rm cont}c_{\rm f})]$ where $S$ is the flux of the absorption, $S_{\rm cont}$ of the continuum and $c_{\rm f}$ the covering factor which we assume to be 1), would range between $\tau = 0.006$ (first scenario, circumnuclear disk) and $\tau = 0.018$ (second scenario, radio bubble interacting with the ISM).
The corresponding column density of the \HI\ outflow ranges between  $5 \times 10^{18} T_{\rm spin}$ cm$^{-2}$ for  the case  the absorption being against the disk (with  \tspin\ being the spin temperature of the \HI), to $1.5 \times 10^{19}  T_{\rm spin}$ cm$^{-2}$ if the absorption is against the inner nuclear bubble structure. 

The  correct value to use for  \tspin\ is quite uncertain. The typical spin temperature assumed is  the kinetic temperature of the \HI\ which is around 100 K. But this is  under the condition that  the excitation of the gas is not affected by the radiation field of a powerful, nearby continuum source. As described in  \citet{Bahcall69,ODea94} and \citet{Maloney96}, if the absorbing gas is located close to strong source of radiation, e.g.\ an AGN,  the \tspin\ can increase to  thousands of K, depending on the density of the gas.
This is likely to be relevant for the \HI\ gas in Mrk~231, given that we know the absorption is coming from  the nuclear region. 

\begin{figure}
	\centering
\includegraphics[width=9cm,angle=0]{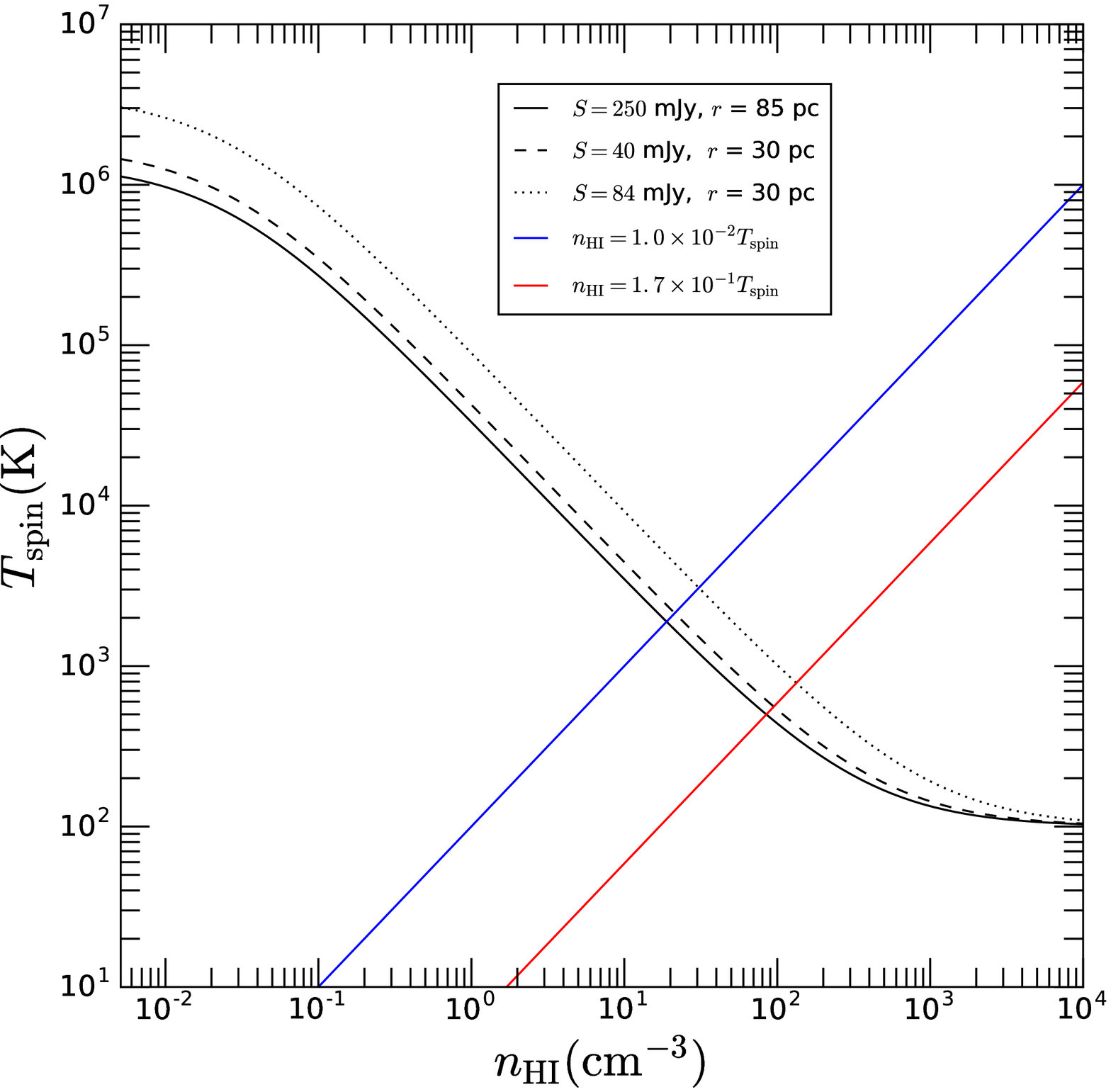}
	\caption{\label{fig:Tspin}  Plot of the \tspin\ as function of the \HI\ density following \citet{Bahcall69} for three different combinations of locations and fluxes as described in Sec.\ \ref{sec:HIabs} (black lines). The red and blue lines indicate the density as function of  \tspin\ as derived from the absoprtion data of Mrk~231. The intersection defines a relatively narrow region for the values of  \tspin\ and density for the outflowing \HI\ (see Sec. \ref{sec:HIabs} for details). }
\end{figure}

For both scenarios discussed above, one can calculate, using the formulae presented in \citet{Bahcall69}, the \tspin\ of the absorbing material as function of the density of the gas, given the assumed location of the gas and the strength of the relevant continuum source and this is given in Fig.\ \ref{fig:Tspin}.
The three black curves show  \tspin\ as function of the assumed density $n_{\rm HI}$  for clouds in the two extreme situations which could be present in Mrk~231: i) the gas is located at an average distance of 85~pc,  i.e.\ the average radius ($r/2$) of the disk, and affected by the full flux in the central region of the source ($\sim 250$ mJy); or ii) the gas is located at the distance of the inner "bubble" (i.e.\ 30 pc) and experiencing the flux from either only the radio core (i.e.\ 40 mJy) or from core plus "bubble" (i.e.\ 84 mJy). Figure \ref{fig:Tspin} shows that only for high densities, \tspin\ is expected to be as low as 100 K.

Which part of Fig.\ \ref{fig:Tspin} is relevant for Mrk~231 can be found by using  constraints from the observed optical depth of the \HI\ absorption.
If the \HI\ absorption is due to a structure with depth comparable to the one of the disk  (i.e.\ 170 pc) illuminated by the full flux in the central region, the data give $n_{\rm HI} = 1.0 \times10^{-2} T_{\rm spin}$ cm$^{-3}$. For the other extreme, where the \HI\  is located in a structure with depth comparable to the size of the inner radio structure  (i.e.\ 30 pc), we derive  $n_{\rm HI} = 1.7 \times10^{-1} T_{\rm spin}$ cm$^{-3}$. These relations are indicated in Fig.\ \ref{fig:Tspin} by the  red  and blue lines and the region defined by the intersection of the  lines identifies the range of possible values for \tspin\ and densities. From Fig.\ \ref{fig:Tspin} we see that  \tspin\  ranges between 400 and 2000 K and the corresponding densities are in the range  10 - 100 cm$^{-3}$.
The \tspin\ we obtain is consistent with what was derived by \citet{Carilli98} for the \HI\ in the disk in order to obtain  \HI\ densities consistent with the results on the  free-free absorption \citet{Taylor99}.  The densities we derive for the outflow are somewhat lower those derived the  from free-free absorption (see Sec. \ref{sec:literatureLine}), but this may not be unexpected if the fastest outflowing gas is associated with the lower density parts of the outflow.

We can use these parameters to derive the \HI\ mass outflow rate. 
This can be estimated following \citet{Heckman02} and \citet{Rupke02,Rupke05}. Different assumptions can be made on the  size of the outflowing region, going from 200 mas (about 170 pc; comparable to the radius of the \HI\  central disk ) to 35 mas (about 30 pc)  corresponding to the distance of the radio "bubble" from the core. A opening angle  of $1 \pi$  steradians has been assumed for the outflow and to derive the column density we use   \tspin \ = 1000 K. Using the FWHM of 837 \kms\  of the outflow (see Table \ref{tab:GaussFit}), we find the \HI\ mass outflow rate  to  be between 8.1 and 18.5 \msunyr. The kinetic energy associated with this outflow is between $5 \times 10^{42}$ and $1.1 \times 10^{43}$  \ergs. These values are comparable to what is found for radio galaxies \citep{Morganti05a}. They are lower than what is derived for the mass outflow rate of the molecular gas in Mrk~231 \citep{Feruglio15}, but higher than for the ionised gas.  Below we discuss this in more detail.

%============================ Plot Tspin ======================================

%===========================================================================================

%============================ VLA radio continuum ======================================

\begin{figure}
	\centering
\includegraphics[width=9cm,angle=0]{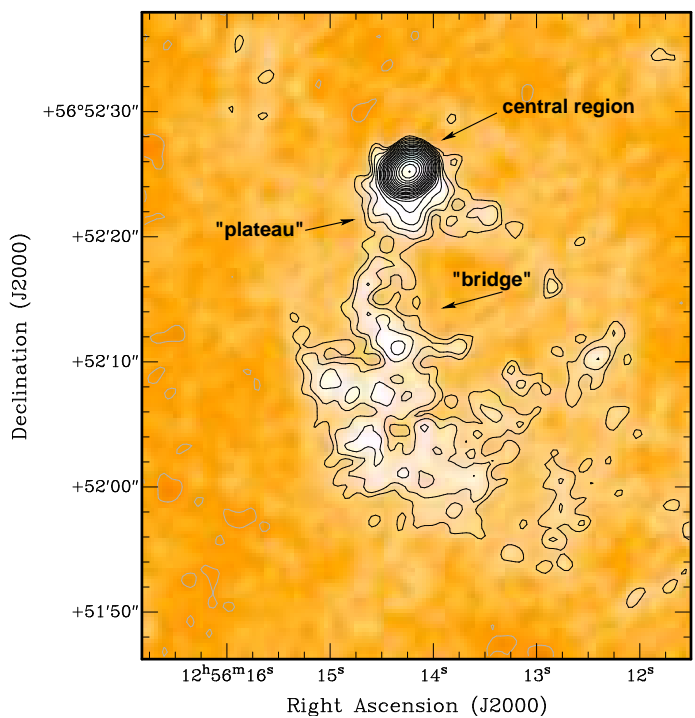}
	\caption{\label{fig:ContRob} Radio continuum image obtained from the VLA data using  robust 0.5 weighting ($1^{\prime\prime} =  0.867$ kpc). The different  components (bright core, "plateau", bridge structure in the southern lobe) are clearly visible. See text for details. The contour levels range from $-0.07$, 0.07 \mJybeam\  to 250 \mJybeam\ with  increasing factors of 1.5.}
\end{figure} 

%===========================================================================================

\subsection{The radio continuum}

The radio continuum images obtained with the WSRT and VLA dataset are shown in Figs \ref{fig:ContWSRT} and  \ref{fig:ContRob} respectively. In the high resolution VLA image, three main structures can be seen: a bright nuclear component, a low surface brightness "plateau" encircling the southern edge of this central component  and  faint emission from the lobe extending $\sim 30^{\prime\prime}$ to the south.  

\begin{figure}
	\centering
%{\includegraphics[width=9cm,angle=0]{fig7a.eps}
{\includegraphics[width=9cm,angle=0]{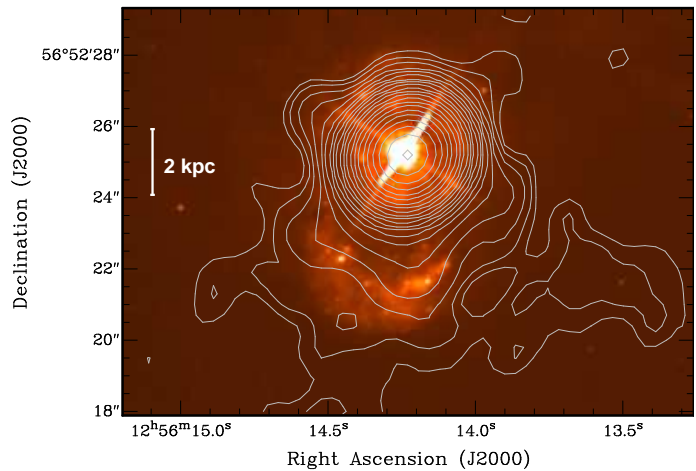}}
	\caption{\label{fig:ContHST} Zoom-in of the central region of the radio source. Radio contours from the VLA image of Mrk~231 are superposed to the false colour HST WFC image made from the F814W (red) and F435W (blue) data.  The  image illustrating the spatial coincidence of the radio plateau  and the optical arc feature shown by HST, representing a region of star formation.}
\end{figure}

From the  uniformly weighted VLA image we find that the bright central structure (with size $< 0.8$ kpc, i.e.\ unresolved by our observations) contains the vast majority of the flux. No jet-like structure is seen extending from this compact nuclear region and the very high contrast between this component and the extended emission is confirmed. The "plateau" extends to about 4 kpc south from the central region and has an integrated flux at 1.3~GHz of only $\sim 1.8$ mJy, with this estimate being quite uncertain.  Interestingly, this structure appears to be coincident with an optical arc seen by HST as illustrated by the overlay in the right panel of  Fig.\ \ref{fig:ContHST}. 

The details of the extended southern structure are seen better than  before and are illustrated in the image shown in Fig.\ \ref{fig:ContRob}. In particular, we trace a faint bridge-like structure, possibly connecting the core to the extended structure. 

\begin{figure}
	\centering
\includegraphics[width=9cm,angle=0]{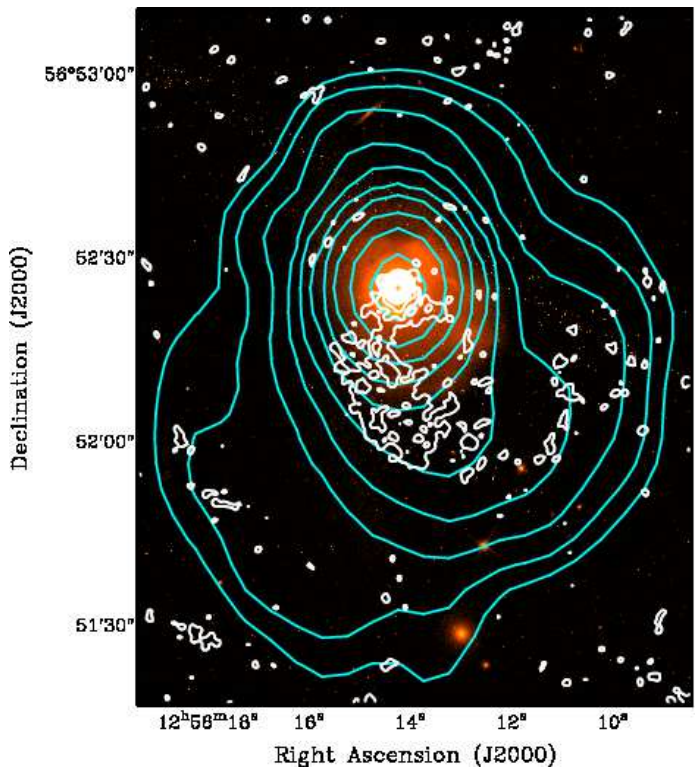}
	\caption{\label{fig:VLA-WSRT} HST image with superimposed the WSRT  (cyan) and the VLA (white) radio contours. The figure illustrate the extension of the large scale structure of the continuum emission recovered by lower spatial resolution of the WSRT image.  Contours like in Fig.\ref{fig:ContRob} and Fig.\ref{fig:ContWSRT}.}
\end{figure}

With the high spatial resolution of the VLA data we recover only partly the diffuse, low surface brightness component. The full extent of the large-scale diffuse lobe is much better recovered by the low spatial resolution of the WSRT image as shown in Fig.\ \ref{fig:VLA-WSRT}. The southern lobe extends about 50\arcsec\ (43 kpc), i.e.\ more than previously recovered.  Interestingly, we find a weak component  extending up to  about $20^{\prime\prime}$ (17 kpc) to the north. 
Thus, in the WSRT continuum image, the structure of the extended radio continuum is less asymmetric than previously thought.

%======================= Overlay VLA radio continuum and HST ================================

%======================================================================================

Deep 500-ksec {\em Chandra} soft (0.5-2 keV) X observations of Mrk 231 \citep{Veilleux14} have revealed a giant (65 $\times$ 50 kpc), hot (several $\times 10^6$ K), and metal-enriched X-ray emitting halo which
shares no resemblance with the tidal debris  seen at optical wavelengths. The {\em Chandra} image is reproduced in Fig.\ \ref{fig:cxo-radio}  where it is compared with the new WSRT and VLA data. While both the hot thermal gas traced by the X-rays and the relativistic non-thermal component
mapped in the radio extend beyond $\sim$30 kpc from the active nucleus, their morphologies are quite different.  The large-scale X-ray emission is somewhat boxy and lopsided to the south-east, while the radio emission detected by WSRT is elongated along the north-south
axis and shows a distinct asymmetry to the south-west (although our new WSRT data now reveals a faint extension to the
south-east albeit on a larger spatial scale than the X-ray emission). A comparison of the X-ray and radio emission on smaller scales ({\em Chandra} vs VLA) also does not reveal any obvious close physical connection between the two phases of material, except perhaps $\sim$15\arcsec\ ($\sim$13 kpc) south of the nucleus where excess radio emission at the intersection of the bridge and southern lobe identified in the VLA map (see Fig.\ \ref{fig:ContRob}) appears to coincide with a minimum in a string of X-ray emission peaks (also seen in Figure 4 of \citealt{Veilleux14}). We speculate that the relativistic material at that location fans out to produce the southern lobe, and curves to the south-west to avoid the denser X-ray halo material on the south-east.

As described in Sec.\ \ref{sec:VLAobs}, the image at 1.6~GHz  suffers from  poor quality of the data (in particular by the presence of strong  RFI). Nevertheless, also at this frequency we detect the emission from the plateau.  It is difficult to derive a reliable spectral index between these close frequencies (1.3 and 1.6 GHz). However, a first order estimate for the   region of the plateau  gives values between  $-0.7$ and $-0.8$ (for $S \sim \nu^{\alpha}$), thus the spectrum is relatively steep. 
Considering the coincidence between the radio emission of the plateau and the arc of optical emission detected by HST  (which is likely associated with a region of star formation), the radio emission  may  originate from star formation. Following  \citet{Condon02} and assuming an integrated flux of 1.8 mJy (i.e.\ a  luminosity of about $7 \times 10^{21}$ W Hz$^{-1}$),  the corresponding star formation rate would be $\sim$5 \msunyr.
In order to compare this value with that derived from  the optical image, the flux in the F435W HST image was integrated over a 13.6 square arcsecond region surrounding the blue arc. This gives a Vega magnitude of 19.1. The original Starburst99 \citep{Leitherer99} tables were then used with a stellar mass range of 0.1-100 \msun\ and a Salpeter Initial Mass Function with $\alpha = 2.35$. We obtain 9 \msunyr\  for a  burst  assumed to be continuous and  observed at 10 Myr. Thus, considering the uncertainties in these assumptions and on the
 radio flux of the plateau, the star formation rate obtained from the optical emission is comparable to that derived from the radio
 emission, adding support to the idea that both the optical and radio features are associated with on-going star formation.

\section{Discussion}

Fast \HI\ outflows with similar properties to the one found in Mrk~231 have been detected in a number of objects. Many of these outflows occur in powerful radio galaxies. This can be the result of an  observational bias because their strong radio background makes it easier to detect absorption at the low optical depth typical of the \HI\  outflowing component  (see e.g., \citealt{Morganti05a,Morganti13}). In the handful of objects where the location of this \HI\ outflow  can be determined,  the outflow is seen off-nucleus, suggesting the radio plasma jet is responsible for driving the outflow. The best examples of this are 3C~305 and 4C12.50 (see \citealt{Morganti05b,Morganti13} respectively).

However,  a number of objects with low radio luminosity  are now known to also  show high-velocity \HI\ outflows \citep[see e.g.,][]{Alatalo11,Shafi15,Oosterloo00}, thus suggesting that \HI\ could be common also in gaseous outflows originating in different conditions.  The less powerful objects represent a mix of situations, where the radio jet clearly plays a major role in some cases (e.g.\ IC~5063, \citealt{Oosterloo00,Morganti15}), while in other cases other processes are probably also at work. 

Detecting the \HI\ outflow in Mrk~231, a well studied object where outflows of many different phases of the gas (from X-ray emitting to cold molecular)  have been already traced, adds a case to this second group. The diversity of objects displaying fast outflows of cold gas suggests that their occurrence depends less on which mechanism is driving the outflow and more
on the environment surrounding the energy source responsible for the outflow. Furthermore, Mrk~231 reinforces the idea that fast outflows are multiphase in nature, a characteristic that should be taken into account when explaining such outflows with theoretical models.
 
Below we discuss in  more detail how the \HI\ outflow compares with outflows already detected in  other phases of the gas in Mrk~231 and what this can tell us about the driving mechanism.

%======================= Overlay WSRT and VLA radio continuum ===============================

%=====================================================================================

\subsection{Comparison with other phases of the gas: \Na} 

It is interesting to compare the characteristics of the \HI\ outflow with the results from \citet{Rupke11,Rupke13a} where they trace the neutral gas outflow using \Na\ using the absorption doublet 5889.95 and 5895.92 \AA\ (the Na D doublet) which can be detected against the stellar light of the host galaxy. \Na\ is a good tracer of cold neutral gas because its ionisation potential (5.1 eV) is lower than that of hydrogen. Thus,  \Na\  and \HI\ absorption trace similar phases of gas under similar conditions in the ISM. Therefore, to first order, we expect the characteristics of these outflows to be similar. \Na\ outflows have been detected before in radio sources and ULIRGs \citep{Rupke02,Rupke05,Lehnert11,Rupke13a,Cazzoli16}.

\citet{Rupke11} and \citet{Rupke13a} have detected a blueshifted component of the \Na\ absorption in Mrk~231  extending in every direction from the nucleus  out to at least 3~kpc. They have explained this as a wide-angle wind.  
Our results on the structure of the radio continuum exclude the presence of a radio jet on the 3-kpc scale, suggesting that the wind is not
driven by the jet.
This is further supported by lack of obvious signatures of strong localized shocks in the X-rays \citep{Veilleux14} expected when a beamed radio source strongly interacts with the surrounding ISM. 

A number of similarities can be seen between \Na\ and \HI\ outflows. The broad blueshifted absorption components of \Na\ cover similar velocities as those we detect  in the \HI\ outflow.  
The  column density of the \HI\  in our data (ranging from 5 to 15$ \times 10^{21}$ cm$^{-2}$ for \tspin = 1000 K) is also consistent with the column density derived for the  \Na\ ($7.5 \times 10^{21}$ cm$^{-2}$). 
All this suggests that the  \Na\ and \HI\ absorption may indeed come from the same outflow. The difference in distribution and extent of the background continuum (stellar light in one case and radio continuum in the other) is the likely explanation for the \Na\ being observed over a larger area, extending to about 3~kpc from the nucleus. 
The mass outflow rate derived by \citet{Rupke13a} is $\sim 179$ \msunyr, higher than we derive   for the \HI. However, this difference is likely due to different  sizes of the outflowing region in the calculation of the mass outflow rate. 
Therefore, we conclude that {\sl \Na\ and \HI\ are likely part of the same outflow}, where the \HI\ traces the inner regions - where the \Na\ outflow cannot be traced because the continuum is completely dominated by the small-scale quasar light and the host galaxy is lost in its glare -  while the \Na\ is telling us about the larger scale - where the \HI\ cannot be traced because no background radio continuum is present.

%====================== Overlay X-ray with WSRT and VLA radio continuum ===============================

\begin{figure}
	\centering
\includegraphics[width=9cm,angle=0]{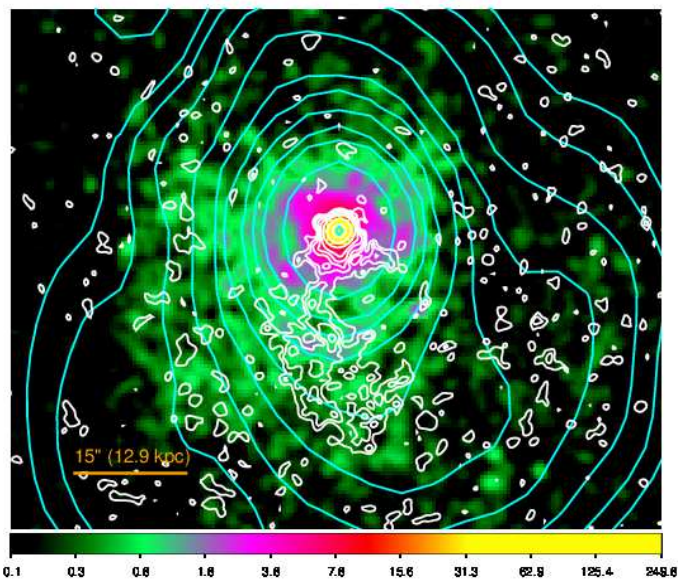}
	\caption{\label{fig:cxo-radio} Chandra soft (0.5-2 keV)
X-ray image from \citep{Veilleux14} with overlaid contours of  the VLA and WSRT continuum images. The X-ray image has been smoothed by a 3 pixel Gaussian to best match the VLA beam. Contours as in Figs \ref{fig:ContRob} and \ref{fig:ContWSRT}.}
\end{figure} 

%=====================================================================================

\subsection{Comparison with other phases of the gas: molecular gas} 

Unlike \HI\ and \Na, the molecular gas in Mrk 231 is observed in both absorption and emission, and therefore can provide some information on
the full extent of the outflow.The study of \citet{Feruglio15} shows that the molecular CO(2-1) outflow extends in all directions around the nucleus, being more prominent along the south-west to north-east direction.   Extended, redshifted emission with lower surface brightness is seen north-east from the nucleus out to $\sim$ 1 kpc. However, the bulk of both receding and approaching outflowing gas is located within $\sim 400$ pc from the nucleus, and peaks $\sim$ 0.2 arcsec south-west of the nucleus.  The highest velocities reached by the molecular outflow are about 1000 \kms, comparable to those observed in \HI.
Similar outflow velocities are seen in OH with Herschel (e.g., \citealt{Fischer10,Sturm11}).

In analogy with the proposed geometry of the neutral outflow
traced by \Na, the outflow  of cold molecular has been explained as a nuclear wind with a wide-angle biconical geometry, as illustrated in Fig.\ 17 of \citep{Feruglio15}. 
The mass outflow rate is 500--1000 \msunyr\ out to $\sim1$ kpc  
 \citep[using a conservative conversion factor CO to H$_2$ of  0.5;][]{Weiss01} and the kinetic power $\dot E = 7$--$10 \times10^{43}$ erg s$^{-1}$ which corresponds to about 1-2\% of the bolometric AGN luminosity.  The impact of such an outflow on the star formation has already been discussed in \citet{Feruglio10}. This is further confirmed by the fact that the energetics derived for the molecular gas by \citet{Feruglio15} agree well, whithin the uncertainties, with those
of \citet{Gonzalez14} and Gonz{\'a}lez-Alfonso et al.\ (in prep.)  derived from the velocity-resolved profiles of multiple OH transitions obtained with Herschel-PACS. 
Thus, the mass outflow rate associated with the molecular gas is much higher than what derived for the \HI. This makes the outflow in Mrk~231 similar to most of the other cases where both atomic and cold molecular outflows are detected where the \HI\  outflow is always much less massive than the molecular one. 

\subsection{Origin of the \HI\ outflow }
\label{sec:origin}

In the case of Mrk~231, even our highest resolution image does not allow us to locate the region of the outflow, thus we cannot {\sl directly}  determine whether the outflow is driven by a wind or a jet.
Here we estimate whether the energetics in the nuclear radio "bubble" would be enough for driving the outflow.
A number of studies have proposed relations between   radio luminosity and jet power \citep{Willott99,Wu09,Cavagnolo10}. Especially in the case of complex objects like Mrk~231, these relations should be taken with care and indeed shortcomings of them have been pointed out (see \citealt{Godfrey16}). 
Nevertheless, we obtain a first order indication of the jet power by using  the relations proposed by \citet{Willott99},\citet{Wu09} and \citet{Cavagnolo10}. We use the flux emitted by the core and the brighter radio bubble  (at 30 pc from the nucleus, and the most likely structure that could produce a shocked cocoon). In this way we derive a range of values for the jet power of $\sim 7 \times 10^{42}$ to $\sim 9 \times 10^{43}$ \ergs.
These values are comparable to the kinetic energy of the \HI\ outflow and lower than (or at most comparable to) that of the molecular component. Thus, unless  unrealistic conditions  are assumed (i.e. extremely high efficiency  and/or a dominant thermal component), the jet power  does not seem large enough to drive and sustain the outflow.

Based on this and on the similarities between the velocities and column densities of the \HI\  and the \Na\ outflow, we conclude that the wide-angle wind is likely the (dominant) mechanism at the origin of the \HI\ outflow, although the action of the radio plasma cannot be completely   ruled out until observations of high enough spatial resolution pinpoint the
location of the outflow
 
Various processes can create a large-scale, wide-angle wind (see e.g.\ discussion in \citealt{Veilleux05}). The one often considered is the possibility of having a wind driven by the inner
radiation-pressure originating in the accretion disk. This wind would then collide with and accelerate the ISM. However, other process can also be considered like a hot thermal wind
(e.g., Compton-heated, \citealt{Begelman85}) colliding with and accelerating the ISM, or  even the pc-scale jet producing an over-pressured cavity from which the wide-angled biconical outflow is produced.
 
Our data do not allow us to distinguish between these processes. However, it is worth mentioning that models describing the winds originating by radiation pressure have also now been expanded to explain the presence the cold, molecular component associated with shocks and other energetic phenomena (see e.g.\ \citealt{King03,Faucher12,Zubovas14,Costa14}). According to these models, the wind, originating very close-in and launched by AGN radiation pressure,  strongly  shocks against the surrounding gas, driving the outflow. These models predict the outflow to be unstable for high-temperature and the mixing of the shocked gas and surrounding medium makes the cooling of this  gas very efficient, forming a  two-phase, outflowing medium, with cold dense molecular clumps mixed with hot tenuous gas. Thus, these models may be particularly relevant for Mrk~231 and the presence of a component of atomic neutral gas in the outflow can provide additional constraint.
If the neutral gas outflow is indeed wind-driven, then our \HI\ observations probe the inner portion of this
wind, while the outer portion is probed by the \Na\ observations.

The larger mass and mass outflow rate detected in molecular gas compared to that of the \HI\ is a recurrent characteristic of all  objects where the outflow has been detected in these two phases of the gas  (with the further recurrent characteristic that the ionised gas involves an even lower mass). It has been proposed that the \HI\ - as well as the warm component of the molecular gas, see \citet{Tadhunter14} - represents a short, intermediate stage in the rapid cooling of the gas while the cold molecular gas represents the final phase.  This trend, also confirmed in the case of Mrk~231,  can provide support for the idea that the shocked gas heats up followed by rapid cooling (instead of remaining cold during the entire process) and could provide constraints on the way the cooling is proceeding, fast but slow enough to observe all phases of the gas present in the outflow. 

Finally it is worth noting that, despite the different driving mechanism, the mass outflow rate of the \HI\ outflow in Mrk~231 is in the same range as those observed in known jet-driven outflows (i.e.\ \citealt{Morganti05b,Morganti05a,Mahony13}). This has potentially the interesting  implication that the conditions that allow the gas to cool and form the atomic and molecular components are the same regardless the different driving mechanisms of the outflow. 
The model that at present seems to explain best the characteristics of jet-driven outflows is presented by \citet{Wagner11} and \citet{Wagner12}. In their simulations, the key component is the presence of a {\em porous} medium with {\em dense clumps} which force the jet to find a complex  path of least resistance. In this way, a cocoon of expanding gas forms, accelerating the clouds  to high velocities and over a wide range of directions, away from the jet axis.
It is, therefore, not the  direct  jet-ISM interaction, but the cocoon of shocked gas that, combined with the clumpiness of the medium,  produces the outflow.  Thus, it may not be actually relevant which mechanism is producing a cocoon of shocked gas.  Indeed, \citet{Wagner13} have explained  fast outflows observed in X-ray by using a similar model where this time the cocoon is inflated by a wind from the circumnuclear corona and  which is strongly interacting  with the inhomogeneous, two-phase ISM consisting of dense clouds embedded in a tenuous, hot medium. Thus, in relation to our results for Mrk~231, we suggest that in addition to the mechanism at the origin of the outflow, the clumpy structure of the medium may play an important role in the observed properties of the outflowing gas, together with the amount and phase of the material in the immediate environment of the outflow.

\subsection{Origin of the radio continuum}
  
The radio continuum imaged by our observations traces emission ranging from the kpc to tens of kpc scales. 
Its complex nature - combination of radio emission from star formation and from the AGN -  was already known from  earlier studies \citep{Carilli98,Taylor99}, but it is further emphasized by the structures found with the new data.

The plateau of faint diffuse emission observed around the southern part of the core has been detected for the first time. This component does not appear to be part of the north-south bridge and, furthermore, has a clear optical and X-ray counterpart \citep{Veilleux14}. If due to star formation, as suggested by the coincidence with the optical arc (see Fig.\ \ref{fig:ContHST}), it would correspond to a SFR of $\sim $5 \msunyr, consistent with what found from the optical counterpart ($\sim $9 \msunyr). 

On the few-kpc scale, we do not detect any jet structure  emerging from the central region. However,  we detect a  poorly collimated structure - a bridge - to the southern lobe (see Fig.\  \ref{fig:ContRob}), broadly aligned north-south with the pc-scale bubbles reported by  \citet{Carilli98} and \citet{Taylor99}. 
The origin of the radio emission in the large radio lobe(s) could be due to the fuelling of fresh electrons by the active nucleus or by radio emission from electrons accelerated in-situ, or by a combination of the two.
The fact that we find a bridge structure may support the first hypothesis. On the other hand, the large drop in radio brightness between the core and the extended region  suggests that the electrons producing the radio emission from the nucleus are decelerated and de-collimated in the initial part due to an interaction with the surrounding  medium. The faint, bridge emission in the southern lobe could represent what is left after such interaction, indicating that the interaction does not manage to destroy completely the flow. 
However, shocks could be generated by the interaction between fast outflowing material, like the outflow wind, and the surrounding medium and the radio emission could  originate from electrons accelerated in-situ by these shocks. If the wind is slightly collimated (e.g., by the small-scale disk), it would shock with the ambient material and produce radio emission which would appear slightly collimated.  Possible, the bridge structure is too narrow and too continuous to fully support this hypothesis. However, the presence of in-situ acceleration is suggested by the apparent spatial coincidence between the region where the bridge further de-collimates  and X-ray emission (Fig.\ \ref{fig:cxo-radio}).

Finally, on even larger scales, the WSRT image shows  a small extension to the north which suggests that the overall emission is less asymmetric than previously thought (Fig.\ \ref{fig:ContWSRT}).  
Considering its low power  and the fact that the radio morphology of the inner lobes is more similar to bubbles, the ejection of the radio plasma is unlikely to be relativistic. Therefore, the asymmetry of the lobes cannot be due to relativistic effects. More likely it is connected with effects of interaction with the ISM and with the halo material (i.e. circumgalactic material, CGM).

\section{Conclusions}

Mrk~231 represents another object where we find a fast ($>1000$ \kms)  \HI\ outflow which constitutes  one component of a multiphase AGN-driven outflow.
The multiphase nature of gaseous outflows has now been seen in a growing number of very different objects: in Seyfert galaxies (e.g.\ NGC~3079, IC~5063, NGC~1068), in radio galaxies (e.g.\ 3C~293, 3C~305, 4C~12.50) and in relatively low luminosity AGN (e.g.\ NGC~1266,  NGC~1433). Mrk~231 is yet a different object to add to this list, classified as radio-quiet quasar, and it represents one of the best examples so far where {\sl all} gas phases are now studied in detail. 

The detection of the blueshifted \HI\  component  is complementary to the  outflow traced by  \Na\ and studied by \citet{Rupke11,Rupke13a}. 
The \HI\ outflow  likely represents the inner part of the broad wind  identified on larger scales in atomic \Na. 
This shows that in the outflow a component of atomic  gas  is  present already in the inner regions of the galaxy. The energetics and the similarities with the characteristics of the \Na\ suggests that the \HI\ is, as is the outflow traced by \Na,  the result of a wide-angle wind seen almost face-on. 
Thus,  unlike many other objects where a fast \HI\ outflow has been detected,  the role of the radio plasma jet in producing the outflow is likely not  dominant in Mrk~231.
Nevertheless,  an interaction between the radio plasma and the rich ISM is likely to occur and may explain some of the decollimation of the flow of the radio plasma and the overall distorted radio continuum morphology. 

The mass outflow rate of the \HI\ outflow is relatively low compared to that of the molecular gas but similar to what is found in other radio galaxies, ranging between 8 and 18 \msunyr.  The mass outflow rate derived for \Na\ is higher than what derived by us for the \HI. However, we argue that this difference is due to a systematic difference in
spatial coverage of the neutral outflow by the \HI\ and \Na\ observations.

While the \HI\ may not carry the bulk of the outflowing gas, it is important to note that the atomic and molecular phases appear, at least when deep enough observations are available, hand in hand. Thus, the co-existence of these different phases (already seen in a number of other objects) allow us to use all of them - in an almost interchangeable way -  as tracers of the presence of outflows. 

Furthermore, the results presented here (as well as those for other objects like NGC~3079, \citealt{Shafi15}) shows that the presence of outflows detected using \HI\ absorption is not necessarily associated with an interaction between the ISM and the radio plasma ejected by the AGN. 
The diversity of objects displaying fast outflows of cold gas suggests that their occurrence depends less on which mechanism is driving the outflow, but more on the  environment surrounding  the energy source responsible for the outflow.

Finally, a number of large surveys of \HI\ absorption are planned in the northern and southern hemisphere using radio telescopes equipped with large field of view receivers (e.g.\ ASKAP and Apertif). These surveys will offer the unique opportunity to make a blind inventory of the occurrence of fast outflows of cold gas even in relatively weak radio AGN and will provide targets for follow up at other wavebands to fully quantify their properties.

\begin{acknowledgements}
We thank Jim Ulvestad, Chris Carilli and Greg Taylor for the permission to reproduce their figures. The Westerbork Synthesis Radio Telescope is operated by ASTRON (Netherlands Institute for Radio Astronomy) with support from the Netherlands Foundation for Scientific Research (NWO).  The National Radio Astronomy Observatory is a facility of the National Science Foundation operated under cooperative agreement by Associated
Universities, Inc.
RM gratefully acknowledge support from the European Research Council under the European Union's Seventh Framework Programme (FP/2007-2013) /ERC Advanced Grant RADIOLIFE-320745. S.V. acknowledges partial
support from NASA through grant NNX16AF24G.

\end{acknowledgements}

\end{document}